\documentstyle[11pt,aaspp4,psfig]{article}
\begin{document}

\title{ON THE ORIGIN OF QUASI-PERIODIC OSCILLATIONS AND BROAD-BAND
NOISE IN ACCRETING NEUTRON STARS AND BLACK HOLES}

\author{Dimitrios Psaltis\altaffilmark{1}}

\affil{Harvard-Smithsonian Center for Astrophysics,
60 Garden St., Cambridge, MA 02138;\\ dpsaltis@cfa.harvard.edu}

\author{Colin Norman\altaffilmark{1}$^,$\altaffilmark{2}}

\affil{Department of Physics and Astronomy, The John Hopkins
  University, Charles and 34th Street, Bloomberg Center, Baltimore, MD
  21218;\\norman@stsci.edu}

\altaffiltext{1}{Institute for Theoretical Physics, University of
  California, Santa Barbara, CA 93106}

\altaffiltext{2}{Space Telescope Science Institute, 3700 San Martin
  Drive, Baltimore, MD 21218}

\begin{abstract}
  Accreting neutron stars and black holes share a number of rapid
  variability characteristics, such as quasi-periodic oscillations and
  broad-band noise.  The frequencies of these features were recently
  shown to be comparable to fundamental general relativistic
  frequencies of perturbed test-particle orbits near the compact
  objects. In this paper we propose a dynamical model for producing
  modulations in the inner disk properties at these characteristic
  frequencies.  We postulate that a transition radius exists in the
  accretion disk close to the compact object and investigate its
  response to a broad spectrum of density fluctuations. We find that
  the transition radius acts as a low band-pass filter with a
  broadband response that is constant at frequencies lower than the
  inverse radial drift timescale and decreases above it as the inverse
  of the frequency. Moreover, the response shows strong resonances at
  frequencies nearly equal to the orbital, periastron-precession, and
  nodal-precession frequencies, thereby accounting for the observed
  power-density spectra of accreting compact objects and the
  correlations between their characteristic frequencies.
\end{abstract}

\begin{center}
 Submitted for publication in the {\em Astrophysical Journal}.  
\end{center}
  
\newpage

\section{INTRODUCTION}

Weakly-magnetic accreting neutron stars and black holes are known to
exhibit a variety of phenomena on timescales that span a very wide
range, from days to milliseconds. Soon after their discovery, it
became clear that these objects show large brightness fluctuations as
well as flickering or shot noise (see, e.g., Oda et al.\ 1971). It
was, however, the launch of X-ray timing satellites (such as {\em
  EXOSAT}, {\em GINGA}, and especially {\em RXTE\/}) that led to the
discovery of quasi-periodic oscillations in their X-ray brightness,
with frequencies $\sim 10^{-1}-10^3$~Hz (see van der Klis 1998 for a
recent review).

Quasi-periodic oscillations (QPOs) are a common feature of the
power-density spectra of accreting compact objects and come in many
shapes and flavors. In weakly-magnetic accreting neutron stars, at
least four distinct classes of QPOs have been identified so far: the
$\sim 6-15$~Hz normal branch oscillation (NBO; Middleditch, \&
Priedhorsky 1986), the $\sim 10-70$~Hz horizontal branch oscillation
(HBO; van der Klis et al.\ 1985), and the $\sim 100-1200$~Hz lower-
and upper-kHz QPOs that typically occur in a pair (van der Klis et al.\ 
1996). In black-hole systems, the QPO phenomenology has not been
developed to the same extent so far, mostly because the majority of
black-hole binaries are transient systems and, therefore, not always
observable.  In most canonical spectral states, however, low-frequency
($\sim 10^{-1}-10$~Hz) QPOs and broader peaked noise components (at
$\sim 1-100$~Hz) with varying centroid frequencies have been detected
(see van der Klis 1995 for a review).  Moreover, in microquasars, QPOs
with even higher (up to $\simeq 300$~Hz) and possibly stationary
frequencies are often observed (see, e.g., Morgan, Remillard, \&
Greiner\markcite{MRG97} 1997).

The physical mechanism that produces any (or all) of the rapid
variability components in accreting compact objects is still the
subject of active research. The detection of nearly coherent
oscillations during Type~I X-ray bursts in four low-luminosity neutron
stars with frequencies that are constant in the tails of bursts that
are separated by $\sim$20 months provides an important clue towards
understanding the origin of the kHz QPOs (see Strohmayer et al.\ 
1996). These oscillations are thought to occur at the stellar spin
frequencies, as the latter are coherent and stable to the degree
originally inferred observationally for the burst oscillations. This
property motivated the beat-frequency interpretation of the kHz QPOs
(Strohmayer et al.\ 1996; Miller, Lamb, \& Psaltis 1998), in which the
upper kHz QPO occurs at the Keplerian orbital frequency in a
characteristic radius in the inner disk, whereas the lower kHz QPO
occurs at the beat of the orbital frequency with the stellar spin
(note here that beat-frequency models were originally developed for
the HBO; see Alpar \& Shaham 1985; Lamb et al.\ 1985).

Subsequent detailed analyses of the properties of the burst
oscillations and the kHz QPOs have challenged the beat-frequency
interpretation (see, however, Miller 2000). The frequencies and
amplitudes of the burst oscillations were observed to vary
non-monotonically during the bursts (Miller 1999; Strohmayer 1999).
The peak separation of the kHz QPOs was shown to decrease with
increasing kHz QPO frequencies (van der Klis et al.\ 1997; M\'endez et
al.\ 1998; Psaltis et al.\ 1998) and be statistically different than
the burst oscillation frequencies (M\'endez \& van der Klis 1999).
Moreover, the frequencies of QPOs (and noise components) in both
neutron-star and black-hole systems follow similar tight correlations,
suggesting that all these variability components cannot directly be
related to the presence of a compact object with a hard surface or
large-scale magnetic field (Wijnands \& van der Klis 1999; Psaltis,
Belloni, \& van der Klis 1999; see also van der Klis 1994a, 1994b).

The frequencies of the HBO and kHz QPOs in neutron stars, as well as
of the corresponding QPOs and peaked noise components in black holes,
were recently shown to be nearly equal to the orbital,
periastron-precession, and twice the nodal-precession frequencies of
tilted test-particle orbits at the inner accretion disks (Stella,
Vietri, \& Morsink 1999; see also Stella \& Vietri 1997, 1998, 1999).
These frequencies depend entirely on the gravitational field around
the compact object and not on the accretion flow properties and can,
therefore, account naturally for the tight correlations between the
observed QPO frequencies and their common occurrence in both types of
accreting compact objects.

Modeling theoretically the time-dependence of inner accretion disks
and their emission is challenging and is hampered mostly by our lack
of understanding of the viscous mechanism for angular momentum
transport.  In the case of modeling high-frequency QPOs, the problem
becomes even more difficult because radiation forces and general
relativistic effects are non negligible (see, e.g., Miller et al.\ 
1998). Studies of the inner disk normal modes have been very fruitful
in revealing the potential of trapping low-frequency oscillations very
close to the compact objects (see, e.g., Kato 1989; Nowak \& Wagoner
1991; Perez et al.\ 1997). Some of the resulting diskoseismic modes
have frequencies comparable to the orbital, periastron-precession, and
nodal-precession frequencies of tilted test-particle orbits at the
inner accretion disks. Exciting or trapping these oscillations at a
characteristic radius in the inner disk is therefore a prime candidate
for producing the modulations required in the QPO interpretation
proposed by Stella \& Vietri (1998, 1999).

In this paper we explore an alternative mechanism for producing
high-amplitude variability predominantly at these characteristic
general relativistic frequencies in the inner accretion disks. We are
motivated by the similarity between the observed power-density spectra
of accreting compact objects and the response function to external
perturbations of driven mechanical and electrical systems (such as a
pendulum in a viscous medium or an RLC circuit; see Bendat \& Piersol
1971, \S2.4). We do not require modes with only particular frequencies
to be excited or trapped at a characteristic radius, but rather study
the response of the accretion disk inside this radius to a broad
spectrum of perturbations imposed outside of it. The broad-band noise
component of the power-density spectrum of the compact object is then
dictated by the broad-band frequency response of the inner disk.
Resonances occur at the normal modes of the accretion disk with a
frequency response that is less suppressed than in the nearby frequencies
and, therefore, appear as the QPO peaks in the power-density spectra.
In this picture, both the break frequency of the broad-band noise
component and the QPO frequencies are associated with properties in
the same narrow annulus in the inner accretion disk and therefore the
correlations between them can be naturally accounted for.

\section{ASSUMPTIONS}

The observed QPOs are typically narrow, with fractional widths as low
as $\delta\nu/\nu\sim 10^{-2}$ (see, e.g., van der Klis et al.\ 1997).
Because all characteristic frequencies in an accretion flow have a
strong dependence on radius and height above the equatorial plane,
such small fractional widths severely constrain the location of the
physical mechanism that determines the QPO frequencies, i.e., it
should be associated with a narrow annulus in a geometrically thin
accretion disk, so that $\delta r/r\lesssim \delta\nu/\nu \sim
10^{-2}$ and $h/r\lesssim (\delta\nu/\nu)^{1/2}\sim 10^{-1}$.  This
narrow annulus represents a transition in the accretion disk
properties but its exact nature is not important for the current
study. Hereafter, for simplicity, we will refer to the flow outside
the transition radius as the outer accretion disk and to the flow
inside as the inner accretion disk, even though the transition radius
can be as small as a few Schwarzchild radii.

We describe the structure of the accretion disk using the continuity
equation for the density $\rho$,
 \begin{equation}
 \frac{\partial \rho}{\partial t} +\frac{\partial}{r\partial r}
 (r\rho u_r) + \frac{\partial}{r\partial\phi} (\rho u_\phi) +
 \frac{\partial}{\partial z}(\rho u_z)=0\;,
 \label{cont}
 \end{equation}
and Euler's equations for the three components of the velocity field,
$(u_r,u_\phi,u_z)$, in a cylindrical reference frame,
\begin{eqnarray}
  & &\frac{\partial u_r}{\partial t} 
  +u_r \frac{\partial u_r}{\partial r}
  +u_\phi\frac{\partial u_r}{r\partial \phi} 
  +u_z \frac{\partial u_r}{\partial z}
  -\frac{u_\phi^2}{r}  = -\frac{\partial \psi}{\partial r}
  -\frac{1}{\rho}\frac{\partial P}{\partial r}+N_r\;,
  \label{ur}\\
  & &\frac{\partial u_\phi}{\partial t} 
  +u_r \frac{\partial u_\phi}{\partial r}
  +u_\phi\frac{\partial u_\phi}{r\partial \phi} 
  +u_z \frac{\partial u_\phi}{\partial z}
  +\frac{u_r u_\phi}{r}  =  
  -\frac{1}{\rho}\frac{\partial P}{r\partial \phi}+N_\phi\;,
  \label{uphi}\\
  & &\frac{\partial u_z}{\partial t} 
  +u_r \frac{\partial u_z}{\partial r}
  +u_\phi\frac{\partial u_z}{r\partial \phi} 
  +u_z \frac{\partial u_z}{\partial z}
  =  -\frac{\partial \psi}{\partial z}
  -\frac{1}{\rho}\frac{\partial P}{\partial z}+N_z\;,
  \label{uz}
\end{eqnarray}
where $P=\rho c_{\rm s}^2$, $c_{\rm s}$ is the local speed of sound,
and its value can be determined by solving a local energy equation.
The structure and response of the accretion disk depends on the
gravitational potential $\psi$, the viscous torque described by the
vector $(N_r,N_\phi,N_z)$, which is the projection of the viscous
tensor on the coordinate vectors, and the local heating and cooling of
the accretion flow. We describe now our assumptions regarding these
properties of the central object and the accreting gas.

We perform our analysis neglecting any explicit special or general
relativistic effects. We specify the gravitational potential of the
central object through the relations
\begin{eqnarray} 
  \Omega^2&=&\frac{1}{r}
  \left(\frac{\partial \psi}{\partial r}\right)_{z=0}\;,\\
  \kappa^2&=&2\Omega\left(2\Omega+r\frac{d\Omega}{dr}\right)\;,\\
  \Omega_\perp^2&=&\frac{1}{z}
  \left(\frac{\partial \psi}{\partial z}\right)_{r}\;, 
\end{eqnarray}
where $\Omega$ is the Keplerian orbital frequency, $\kappa$ is the
epicyclic frequency, and $\Omega_\perp$ is the frequency of vertical
oscillations in an annulus in the accretion disk.  In Newtonian
dynamics, $\Omega_\perp=\kappa=\Omega$ for a non-rotating central
object.  However, we allow here for the possibility that the three
frequencies do not satisfy these relations, taking thus into account,
in an approximate way, the general-relativistic effects of periastron
precession and frame dragging (see, e.g., Misner, Thorne, \& Wheeler
1973). This is equivalent with assuming a pseudo-potential for the
gravitational field for which the three characteristic frequencies
have the same value as the exact general relativistic frequencies.

We study the response to perturbations of an annulus in a
geometrically thin accretion disk that is localized in the radial and
vertical directions. We therefore assume that the $r-$ and $z-$ components
of the vector $\vec{N}$ are negligible and set
\begin{equation}
 N_\phi \equiv -u_\phi\Omega_{\rm d}\;,
 \label{visc}
\end{equation}
where $\Omega_{\rm d}$ is defined in this way to be the inverse of the
radial drift timescale. Equation~(\ref{visc}) allows us to avoid
specifying any particular recipe for the largely unknown mechanism
that transports angular momentum in the disk.

Finally, for simplicity, we assume that the perturbations do not alter
the local temperature, and hence the sound speed, in the accretion
flow. Strictly speaking, this assumption is justified for fluctuations
on timescales much shorter than the thermal timescale but breaks down
for slower phenomena. However, our results do not depend strongly on
this approximation, as it will become obvious later. Indeed, previous
studies of isothermal or adiabatic diskoseismic modes produced
qualitatively similar results (see, e.g., Nowak \& Wagoner 1991).

\section{THE RESPONSE OF AN ACCRETION DISK TO A SPECTRUM OF PERTURBATIONS}

As discussed in \S2, we assume that the properties in the accretion
disk exhibit a transition over an annulus of width $\delta r$ at a
characteristic radius $r$. Defining radial averages over the annulus
width of all physical quantities as
\begin{equation}
  \langle f\rangle=\frac{1}{r\delta r}\int_r^{r+\delta r} rfdr,
\end{equation}
 and dropping for clarity the angle brackets, the continuity equation
becomes
\begin{equation}
  \frac{\partial \rho}{\partial t} + 
  \frac{\partial}{r\partial\phi} (\rho u_\phi) +
  \frac{\partial}{\partial z}(\rho u_z)=
  \frac{1}{r\delta r}\int_r^{r+\delta r} \frac{\partial}{\partial r}
  (r\rho u_r)dr=-
  \frac{1}{r\delta r}\left(
    \left[r \rho u_r\right]_{r+\delta r} - 
    \left[r \rho u_r\right]_r\right)
  \label{cont_r}
\end{equation}
The first term in the parenthesis in the right-hand side of
equation~(\ref{cont_r}) corresponds to the incoming mass flux from
the outer radius of the annulus, whereas the second term corresponds
to the outgoing mass flux from the inner radius. We set $[r\rho
u_r]_{r+\delta r}-[r\rho u_r]_{r} \sim -r^2(\rho_{\rm
  in}-\rho)\Omega_{\rm d}$, where $\rho_{\rm in}$ describes the
density at the outer edge of the annulus. The continuity equation
then becomes
\begin{equation}
  \frac{\partial \rho}{\partial t} + \frac{r}{\delta r} \Omega_{\rm d}\rho + 
  \frac{\partial}{r\partial\phi} (\rho u_\phi) +
  \frac{\partial}{\partial z}(\rho u_z)=\frac{r}{\delta r}
  \rho_{\rm in} \Omega_{\rm d}\;.
  \label{cont_f}
\end{equation}

We assume that the steady-state disk is geometrically thin and in
vertical hydrostatic equilibrium, with $u_\phi>> u_r, u_z$. We also
assume that, in steady state, $u_\phi=\Omega r$.  We introduce small
perturbations to all physical quantities (denoted by the superscript
`1') and linearize equations~(\ref{ur})--(\ref{uz}) and (\ref{cont_f})
keeping only terms up to first order in all small quantities. The
resulting linearized equations are
 \begin{eqnarray}
\left(\frac{\partial }{\partial t} 
   + \Omega\frac{\partial }{\partial \phi}
   + \frac{r}{\delta r}\Omega_{\rm d}\right) \rho^1 
   + \frac{\rho}{r} \frac{\partial u_\phi^1}{\partial \phi}
   +\frac{\partial}{\partial z}(\rho u_z^1)
   &=& \frac{r}{\delta r} \rho_{\rm in}^1 \Omega_{\rm d}\;, \\
\left(\frac{\partial}{\partial t} 
   + \Omega \frac{\partial}{\partial \phi}\right)u_r^1 - 2\Omega u_\phi^1
   &=& -\frac{c_{\rm s}^2}{r}
   \left(\frac{\delta r}{r}\right)
   \left(\frac{\rho_{\rm in}^1-\rho^1}{\rho}\right)\;,\\
\left(\frac{\partial}{\partial t}
   +\Omega\frac{\partial}{\partial \phi}
   +\Omega_{\rm d}\right)u_\phi^1
   +\frac{\kappa^2}{2\Omega}u_r^1
   & = & -\frac{c_{\rm s}^2}{r} \frac{\partial}{\partial \phi}
   \left(\frac{\rho^1}{\rho}\right)\;,\\
\left(\frac{\partial}{\partial t}
   +\Omega\frac{\partial}{\partial\phi}\right)u_z^1
   & = & -c_{\rm s}^2 \frac{\partial}{\partial z}
      \left(\frac{\rho^1}{\rho}\right)\;.
\end{eqnarray}
 
Because the above system of equations is linear, the solution for the
perturbed density $\rho^1$ can be written as the convolution
\begin{equation}
  \rho^1(t,\phi,z) = \int_{-\infty}^{\infty} h(\tau; \phi, z) \rho^1_{\rm in} 
    (t-\tau,\phi,z) dt\;,
\end{equation}
where the function $h(\tau)$ is called the `weighting function' of the
linear system. For the linear system to be physically realizable, it
is necessary that it responds only to past inputs (Bendat \& Piersol
1971, \S2.3), and hence that
\begin{equation}
  h(\tau<0) = 0\;.
  \label{hcond}
\end{equation}
Instead of solving for the weighting function $h(\tau)$, we define the
response function $A(\omega)$ as
\begin{equation}
  A(\omega) \equiv \int_{\omega=0}^{\infty} h(\tau) \exp(-i\omega\tau) d\tau
  \label{Af}
\end{equation}
such that the Fourier transform $\tilde{\rho}^1(\omega)$ of the
solution is related to the Fourier transform $\tilde{\rho}^1_{\rm
  in}(\omega)$ of the driving term by
\begin{equation}
  \tilde{\rho}^1(\omega) = A(\omega) \tilde{\rho}^1_{\rm in} (\omega)\;.
  \label{ft}
\end{equation}
Note that in the definition~(\ref{Af}) the lower limit of the integral
is at $\omega=0$, because of condition~(\ref{hcond}).

We can obtain the complete solution of the linear system of equations
by expanding the temporal and $\phi$-dependence of the perturber in
modes of the form
\begin{equation}
  \rho_{\rm in}^1(t,\phi,z)= 
  \sum_m \int_\omega \rho_{{\rm in},m}(\omega;z) e^{-i(\omega t-m\phi)}\;.
\end{equation}
Because $\vert A(-\omega,-m)\vert = \vert A(\omega,m)\vert$ (see also
Bendat \& Pierson 1971), we will consider only the response of the
system to perturbations with positive frequencies. Moreover, we will
consider only positive values for $m$, as negative values correspond
to perturbations that rotate in the retrograde sense with respect to
the local Keplerian velocity and are expected to have weak intrinsic
amplitudes. Setting $\xi\equiv \rho^1/\rho$ and $\xi_{{\rm
    in},m}\equiv \rho_{{\rm in,m}}/\rho$ we obtain
\begin{eqnarray}
  -i(\omega-m\Omega)\xi+\frac{r}{\delta r}\Omega_{\rm d}\xi
  + i\frac{m}{r} u_\phi^1
  -\frac{\Omega_\perp^2}{c_{\rm s}^2}z u_z^1 +
  \frac{\partial u_z^1}{\partial z} 
  & = & \frac{r}{\delta r}\Omega_{\rm d} \xi_{{\rm in},m}\\
  -i(\omega-m\Omega)u_r^1-2\Omega u_\phi^1 & = & 
  \frac{c_{\rm s}^2}{r}\left(\frac{\delta r}{r}\right)
  \left(\xi_{\rm in,m}-\xi\right)\\
  -i(\omega-m\Omega)u_\phi^1+\Omega_{\rm d} u_\phi^1 +
  \frac{\kappa^2}{2\Omega}u_r^1&=&-\frac{1}{r}m c_{\rm s}^2 i \xi\\
  -i(\omega-m\Omega)u_z^1&=&-c_{\rm s}^2\frac{\partial \xi}{\partial z}\;.
\end{eqnarray}
Combining the above into a single equation for $\xi$ gives
\begin{eqnarray}
  \frac{r}{\delta r}\Omega_{\rm d}
  \left(\frac{\xi}{\xi_{{\rm in},m}}\right)^{-1} &=&
  \frac{i}{\omega-\omega_m}
  \left(\Omega_\perp^2 z\frac{\partial\xi}{\xi\partial z}-
    c_{\rm s}^2\frac{\partial^2 \xi}{\xi\partial z^2}\right)
  +\left[-i(\omega-\omega_m)+\frac{r}{\delta r}\Omega_{\rm d} 
  \right] \nonumber\\ 
  & & \qquad
  +\frac{m^2\Omega_{\rm c}^2 (\omega-\omega_m)}
  {-i(\omega-\omega_{p-})(\omega-\omega_{p+})+\Omega_{\rm d}
    (\omega-\omega_m)}\;,
  \label{xieq1}
\end{eqnarray}
where $\omega_m\equiv m\Omega$, $\omega_{p\pm}\equiv
m\Omega\pm\kappa$, $\Omega_{\rm c}\equiv c_{\rm s}/r$, and we have
neglected for simplicity the radial pressure forces..

In principle, equation~(\ref{xieq1}) must be solved for the functional
dependence of $\xi$ on $z$ by imposing proper boundary conditions at
the equatorial plane and at infinity. In previous studies of trapped
oscillatory modes in accretion disks (see, e.g., Okazaki, Kato, \&
Fukue 1987), it was shown that the $z-$dependence of $\xi$ can be
written in terms of Hermite polynomials, each order of which represent
a different vertical mode. Driven by this result, we will simply study
here the response of the disk to modes for which
 \begin{equation}
 \frac{\partial \xi}{\partial z}=n^2\frac{\xi}{z}\;;\quad n=0,1\;,
 \end{equation}
where $n$ represents the vertical mode number. This is equivalent with
retaining only the Hermite polynomials of the lowest two orders in the
expansion of the $z-$dependence of $\xi$. Equation~(\ref{xieq1}) then 
becomes
\begin{eqnarray}
  \frac{r}{\delta r}\Omega_{\rm d}
  \left(\frac{\xi}{\xi_{{\rm in},m}}\right)^{-1} &=&
  \frac{-i(\omega-\omega_{n+})(\omega-\omega_{n-})+(r/\delta r)\Omega_{\rm d} 
    (\omega-\omega_m)}{\omega-\omega_m} 
  \nonumber\\ 
  & & \qquad
  +\frac{m^2\Omega_{\rm c}^2(\omega-\omega_m)}
  {-i(\omega-\omega_{p+})(\omega-\omega_{p-})+\Omega_{\rm d}
    (\omega-\omega_m)}\;,
  \label{xieq}
\end{eqnarray}
where $\omega_{n\pm}\equiv m\Omega\pm n\Omega_\perp$.  The response
function of the transition annulus in the accretion disk to external
perturbations of a given mode is then simply (see, eq.[\ref{ft}])
\begin{equation}
  A_{\rm mn}(\omega)=\frac{\xi}{\xi_{{\rm in},m}}\;.
  \label{resp}
\end{equation}

In the above analysis, we have made the explicit assumption that all
perturbations are isothermal. Therefore, by construction, the
radiation emerging from the transition annulus in the accretion disk
will be unperturbed and therefore time-independent. However, the mass
flux to the inner disk and possibly the loading of the hot
Comptonizing medium responsible for the hard X-ray spectral component
will be modulated with a response function of similar qualitative
characteristics as equation~(\ref{resp}). Modulation of the scattering
optical depth in the Comptonizing medium can produce the strong
photon-energy dependence of the observed modulation amplitudes (see,
e.g., Lee \& Miller 1998; Miller, Lamb, \& Psaltis 1998). We,
therefore, make the working hypothesis that the function~(\ref{resp})
describes also the response of the radiation emerging from the inner
accretion flow to perturbations produced outside the transition
radius.

\subsection{Mode Analysis}

In this section we study in more detail the response
function~(\ref{resp}) derived in \S3 for the three lowest-order modes
of perturbation, $(m=0,n=0)$, $(m=1,n=0)$, and $(m=1,n=1)$. The
response of the inner accretion disk described by these three modes is
shown in Figure~1, for typical values of the characteristic
frequencies.

\begin{figure}[t]
 \centerline{ 
\psfig{file=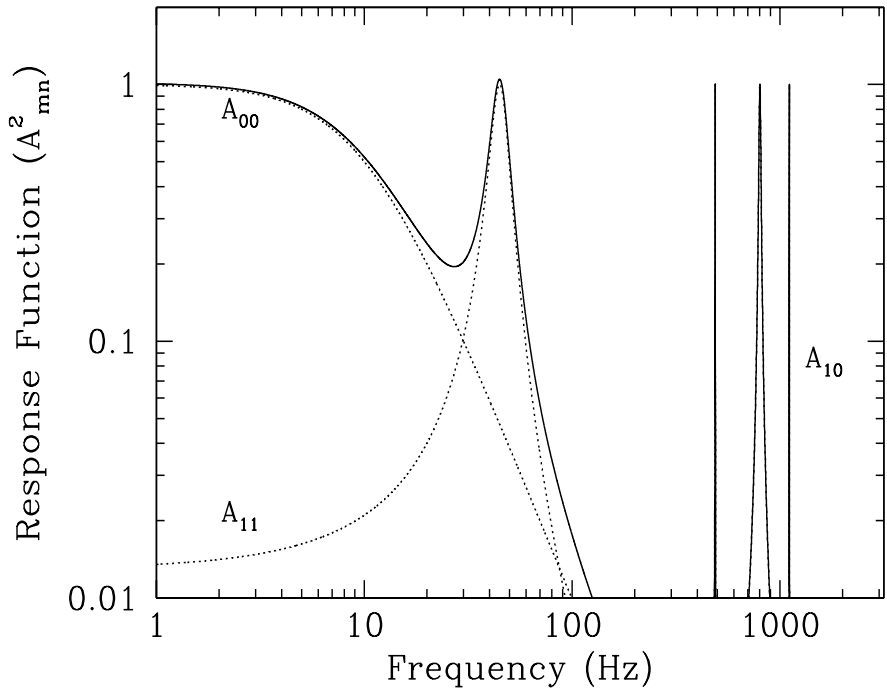,angle=-90,height=10truecm,width=12truecm}}
\figcaption[]{\footnotesize The square of the disk response
  function~(\ref{resp}) for the three lowest-order modes. For the case
  depicted here, $\Omega/2\pi=800$~Hz, $\kappa/2\pi=300$~Hz,
  $\Omega-\Omega_\perp=2\pi 50$~Hz, $\Omega_{\rm d}/2\pi=0.5$~Hz, $\Omega_{\rm
    c}/2\pi=80$~Hz, and $\delta r/r=0.05$.}
\end{figure}

\subsubsection{Mode $(m=0,n=0)$}

This mode describes time-dependent perturbations of the gas
density in the accretion disk that are uniform in both the azimuthal
and vertical directions. The magnitude of its response function 
is
 \begin{equation}
A_{00} = \frac{1}{\left[1+(\omega/\omega_{\rm b})^2\right]^{1/2}}\;,
 \label{A00}
 \end{equation}
where $\omega_{\rm b}\equiv (r/\delta r)\Omega_{\rm d}$. It describes
a low band-pass filter with a response that cuts off at $\omega_{\rm
b}$ and decreases as $A_{00}\sim \omega^{-1}$ at higher
frequencies.

The characteristic properties of $A_{00}$ can be easily understood
given the fact that $\omega_{\rm b}^{-1}$ is the radial drift
timescale in the annulus with radial width $\delta r$. Any density
perturbations with characteristic timescales longer than $\omega_{\rm
  b}^{-1}$ propagate inwards unaffected, whereas perturbations at
increasingly shorter timescales are dumped with an increasingly higher
efficiency.  Note that there is no characteristic frequency in the
disk that corresponds to the $(m=0,n=0)$ mode and hence its response
function shows no resonances.

\subsubsection{Mode $(m=1,n=0)$}

This is an one-armed mode that is uniform in the vertical direction.
As it is also evident in Figure~1, the magnitude of its response is
negligible for most frequencies, besides the frequency ranges near
$\omega\simeq \Omega$, $\omega\simeq \Omega-\kappa$, and
$\omega\simeq\Omega+\kappa$. This is caused by the differential
rotation of gas in the accretion disk that can strongly dump any
azimuthal density perturbations that are driven at frequencies other
than one of the local characteristic frequencies.

Although the complete response is given by the analytic
form~(\ref{resp}), we can obtain some simpler expressions by expanding
relation~(\ref{resp}) around the resonant frequencies. For
$\omega\sim\Omega$,
\begin{equation}
  A_{10}\simeq \left[
    \frac{\omega_{\rm b}^2}
    {\omega_{\rm b}^2+(1+\Omega_{\rm c}^2/\kappa^2)^2
      (\omega-\Omega)^2}\right]^{1/2}\;.
\end{equation}
Therefore, the resonance occurs at the orbital frequency
\begin{equation}
  \omega_3\simeq \Omega
\end{equation}
and the FWHM of the response around $\omega_3$ is
\begin{equation}
  \delta \omega_3\simeq 2\sqrt{3}\omega_{\rm b}
  \left(1+\frac{\Omega_{\rm c}^2}{\kappa^2}\right)^{-1}\;.
  \label{do1}
\end{equation}
 
The response function around the other two resonances is more
complicated and can be simplified analytically in the limit
$\kappa/\omega_{\rm b}\gg 1$, which is usually true for the cases of
interest here. The resonances occur at the frequencies
\begin{equation}
  \omega_{2\pm}\simeq \Omega\pm\kappa
  \left(1+\frac{\Omega_{\rm c}^2}{2\kappa^2}\right)
  \label{o2pm}
\end{equation}
 and the FWHM of the response around these frequencies is
\begin{equation}
  \delta \omega_{2\pm} \simeq \sqrt{3}\omega_{\rm b}
  \left(\frac{\Omega_{\rm c}^2}{\kappa^2}\right)\;.
  \label{do2}
\end{equation}
If, moreover, $(\delta r/r)(\kappa/\Omega_{\rm c})^2<<1$ then the
response function around the resonances can be put in the form
\begin{equation}
  A_{10}\simeq \left\{1+
    \left[\mp \frac{\kappa}{\omega_{\rm b}}+
      \left(\frac{\Omega_{\rm c}}{2\omega_{\rm b}}\right)
      \frac{\Omega_c}{\omega-(\Omega\pm\kappa)}\right]^2\right\}^{-1}\;,
  \qquad\mbox{for}~\omega\simeq \Omega\pm \kappa\;.
\end{equation}

Note that the frequencies $\omega_{2\pm}$ derived here are consistent
with the ones obtained simply by using the dispersion relation of the
inertial-acoustic waves derived by Kato (1989; eq.~[3.13] for $m=1$
and $n=0$)
\begin{equation}
  (\omega-\Omega)^2-\kappa^2=k_r^2c_{\rm s}^2
\end{equation}
and setting their radial wavenumber equal to $k_r\simeq 1/r$. Note,
also, that the amplitude of the response at $\omega_{2\pm}$ is
comparable to unity only in the limit $(\delta r/r)(\kappa/\Omega_{\rm
  c})^2<<1$ and because we have neglected radial pressure forces,
which damp radial oscillations.

\begin{figure}[t]
  \centerline{
    \psfig{file=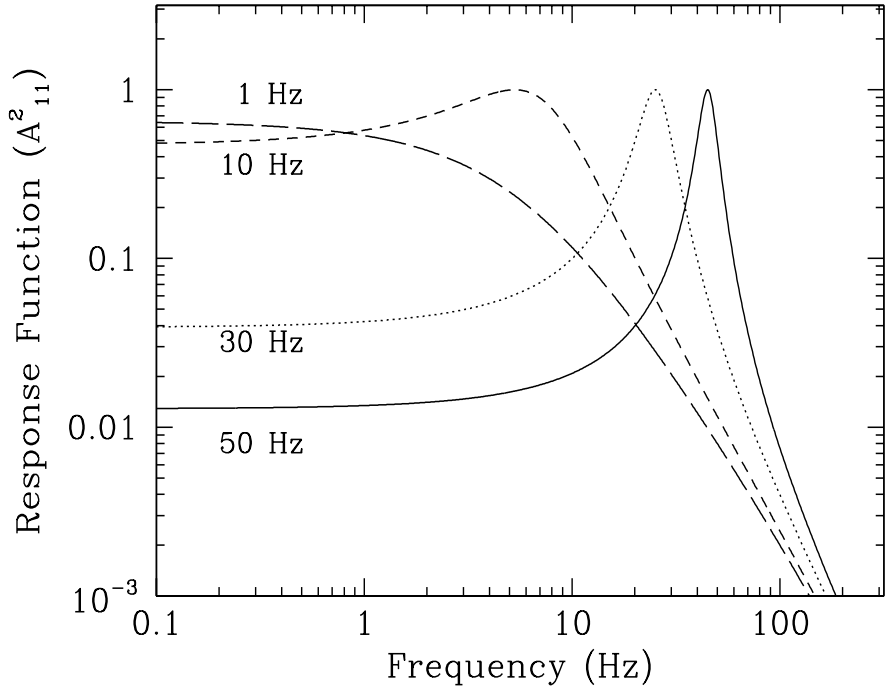,angle=-90,height=10truecm,width=12truecm}}
  \figcaption[]{\footnotesize The square of the disk response function
    for the $(m=1,n=1)$ mode and for different values of
    $(\Omega-\Omega_\perp)/2\pi$. The remaining parameters are the
    same as in Figure~1.}
\end{figure}

\subsubsection{Mode $(m=1,n=1)$}

This mode describes a precessing, one-armed azimuthal density
perturbation with a vertical tilt. As it is also evident in Figure~1,
the differential rotation of tilted orbits in the disk strongly dumps
all such perturbations, unless $\omega\simeq \Omega\pm\Omega_\perp$
(the upper resonance of this mode is not plotted in the figure).

The response function around the precession frequency is
 \begin{equation}
 A_{11}\simeq\left\{
 \left[1+\left(\frac{\delta r}{r}\right)\frac{\Omega_{\rm c}^2
    \Omega_\perp^2}
    {\Omega_{\rm d}^2\Omega_\perp^2+(\kappa^2-\Omega_\perp^2)^2}\right]^2
 +\left(\frac{\omega-\omega_{1\pm}}{\omega_{\rm b}/2}\right)^2\right\}^{-1}
 \;,\qquad\mbox{for}~\omega\simeq\Omega\pm\Omega_\perp\;,
 \end{equation}
 where
 \begin{equation}
 \omega_{1\pm}\equiv \Omega\pm\Omega_\perp\left[1+
 \frac{\Omega_{\rm c}^2(\Omega_\perp^2-\kappa^2)}
 {2\Omega_{\rm d}^2\Omega_\perp^2+2(\kappa^2-\Omega_\perp^2)^2}
 \right]
 \label{o1}
 \end{equation}
 Typically, the response $A_{11}$ peaks at $\omega\simeq
 \omega_{1\pm}$ with a FWHM of
 \begin{equation}
 \delta\omega_{1\pm}\simeq \sqrt{3}\omega_{\rm b}
 \left[1+\left(\frac{\delta r}{r}\right)\frac{\Omega_{\rm c}^2
     \Omega_\perp^2}
  {\Omega_{\rm d}^2\Omega_\perp^2+(\kappa^2-\Omega_\perp^2)^2}\right]\;.
\end{equation}
How pronounced the peak at $\omega_{1-}$ is depends on the relative
magnitude of the radial drift timescale and the resonance frequency
(Fig.~2). As $\omega_{1-}$ decreases, the resonance peak becomes
broader, the low-frequency response for this mode becomes comparable
to unity, and the overall response shows a break at a frequency
comparable to $\omega_{1-}$. For the response function to show a
narrow and pronounced resonance peak, two conditions need to be
satisfied, i.e., $(\delta\omega_{1-}/\omega_{1-}) \ll 1$ and
$\lim_{\omega\rightarrow 0}A_{11} \ll 1$, both of which lead to the
single requirement $\omega_{\rm b}\ll \omega_{1-}$. Note also that the
amplitude of the response at $\omega_{1\pm}$ is unity only because we
neglected $N_z$ and hence the damping of this mode caused by the
differential precession of nearby fluid elements..

\section{APPLICATION TO GALACTIC NEUTRON STARS AND BLACK HOLES}

In this section we study the applicability of the disk modes analyzed
in \S3 in modeling the X-ray variability properties of galactic
neutron stars and black holes. We first address the basic assumptions
in our analysis and then compare directly the predicted mode
frequencies to the observed QPO and noise frequencies in a variety of
sources.

The presence of a sharp transition in the properties of a
geometrically thin accretion disk is a generic requirement of {\em
  any\/} model that attributes any of the observed QPO and noise
frequencies to characteristic frequencies in the disk. The physical
mechanism that produces this sharp transition is not specified in our
analysis. Indeed, the predicted mode frequencies depend mostly on the
gravitational field around the compact object and only weakly on the
hydrodynamic properties of the flow itself. As a result, different
mechanisms may give rise to the sharp transition in different sources
and yet produce QPOs and noise with the same characteristic
frequencies. This is an important property of the model given the fact
that similar QPOs and noise components have been identified in sources
with widely different masses, accretion rates, and magnetic field
strengths, but with centroid frequencies that follow very similar
correlations (see, e.g., Psaltis et al.\ 1999a). Sharp transitions in
accretion disk properties have been discovered in a variety of
situations, such as the transition around the sonic point caused by
radiation drag (Miller et al.\ 1998), the heating front during a dwarf
nova instability (see, e.g., Menou, Hameury, \& Stehle 1999), or the
transition between a thin accretion disk and an advection-dominated
accretion flow (see, e.g., Kato \& Nakamura 1998). Any or all of these
mechanisms might be responsible for the transition that causes the
observed X-ray variability in different accreting compact objects.

In the physical picture presented here, the various disk modes need
not be excited or indefinitely trapped only in a localized region of
the accretion flow. In fact, some of the basic features of the model
depend on the slow leakage of power from the different modes excited
outside the transition radius, which is modeled by the parameter
$\Omega_{\rm d}$ in \S2. This leakage is what generates the broad-band
response plotted in Figure~1 and hence the noise component of the
observed power-density spectra in galactic sources. Moreover, if the
leakage of power is slower than the mode period, it will not affect
the presence of the resonance peaks but will determine only their
widths. This is a general argument regarding the observational
signatures of disk modes, applicable also to the trapped but slowly
leaking modes reported elsewhere (e.g., Okazaki et al.\ 1987; Kato
1989, 1990; Nowak \& Wagoner 1991; Perez et al.\ 1997, etc.). It is,
however, important that significant power is generated close to the
transition radius, because the characteristic timescales at larger
radii are too short for strong perturbations to be generated at
frequencies comparable to the higher-frequency resonance peaks.

Although the small fractional QPO widths imply that their frequencies
are determined in a characteristic radius in the accretion disk, their
hard X-ray spectra (see, e.g., Berger et al.\ 1996) suggest that the
emission from the hot component of the accretion flow is also
modulated at the QPO frequency. If the hard power-law components of
the X-ray spectra are produced by thermal Comptonization of soft
photons by hot electrons, then a small-amplitude oscillation of the
density in the Comptonizing medium can account for the observed QPO
energy spectra (Lee \& Miller 1998; Miller et al.\ 1998). This has
been our motivation for calculating the frequency response of the
density in the inner accretion disk to a broad spectrum of external
perturbations.

We have performed our analysis neglecting any explicit special or
general relativistic effects. However, we have allowed for the
possibility that the azimuthal and vertical frequencies, $\Omega$ and
$\Omega_\perp$, are not equal and that the periastron precession
frequency is not zero. Following Stella et al.\ (1999), we will use
the weak-field and slow-rotation limits of the general relativistic
expressions for these frequencies and compare the predicted
frequencies to the observed QPOs (note here that our expressions are
not strictly valid for a rapidly rotating neutron star, the external
spacetime of which is not described by the Kerr metric). Given a
transition radius $r$, as well as a mass $M$ and specific angular
momentum per unit mass $a_*$ of the central compact object, we obtain
for the corresponding frequencies for a prograde disk, as measured by
a static observed at infinity (see Bardeen et al.\ 1972; Perez et al.\ 
1997; Stella et al.\ 1999),
 \begin{eqnarray}
 \Omega^2&=&\frac{M}{r^3[1+a_*(M/r)^{3/2}]}\;,
 \nonumber\\
 \kappa^2&=&\Omega^2(1-6M/r)\;,
 \end{eqnarray}
and
 \begin{equation}
 \Omega_\perp^2=\Omega^2[1-4 a_*(M/r)^{3/2}]\;,
 \end{equation}
where we have set $G=c=1$, $G$ is the gravitational constant and $c$
is the speed of light. Finally, we estimate the hydrodynamic
timescales related to the accretion flow using the standard
$\alpha$-prescription for the viscosity (Shakura \& Sunyaev 1973; see
also Frank, King, \& Raine 1992), as
\begin{equation}
  \Omega_{\rm V}\simeq \alpha (h/r)^2 \Omega\;,
\end{equation}
and
\begin{equation}
  \Omega_{\rm c} \simeq (h/r) \Omega_\perp\;,
\end{equation}
where $\Omega_{\rm V}^{-1}$ is the viscous timescale, $h$ is the
scale height of the disk, and $\alpha$ is the viscosity parameter,
and set for simplicity the radial drift timescale at the transition
radius equal to the viscous timescale, i.e., $\Omega_{\rm
  d}=\Omega_{\rm V}$.

Following Stella et al.\ (1999) we identify (a) the upper kHz QPO with
the resonance at $\omega_3$ (i.e., the Keplerian frequency) of the
$(m=1,n=0)$ mode, (b) the lower kHz QPO, or peaked noise component
identified as such, with the resonance at $\omega_{2-}$ of the
$(m=1,n=0)$ mode, and (c) the HBO, or low-frequency QPO identified as
such, with {\em twice\/} the frequency of the resonance at $\omega_1$
of the $(m=1,n=1)$ mode; the only reason for choosing twice the
frequency $\omega_1$ is to account for the relatively high observed
HBO frequencies in neutron star systems, which are otherwise
inconsistent with the properties of any stable star in general
relativity (Stella et al.\ 1999; see also Psaltis et al.\ 1999b;
Kalogera \& Psaltis 2000). The latter choice can be motivated by the
presence of strong subharmonics of the QPOs in both black-hole and
neutron-star systems\footnote{We treat here the `intermediate' branch
  of the HBO discussed by Psaltis et al.\ (1999a; see also Ford \& van
  der Klis 1998) as evidence for subharmonic structure.}  and
justified by the two-fold symmetry of the $(m=1,n=1)$ mode.

\begin{figure}[t]
  \centerline{
    \psfig{file=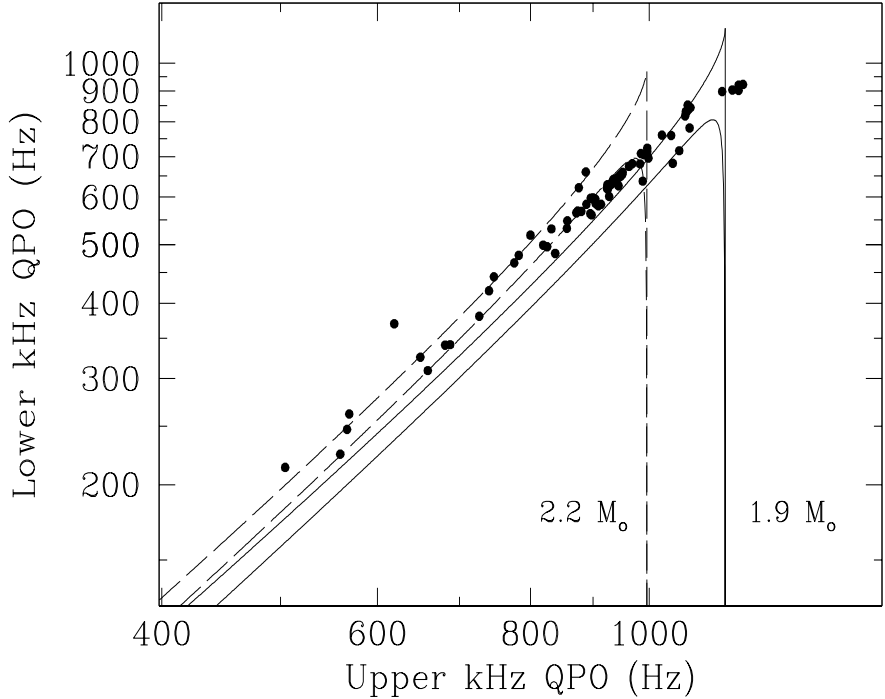,angle=-90,height=10truecm,width=12truecm}}
  \figcaption[]{\footnotesize Correlation between the lower and upper
    kHz QPO frequencies in galactic neutron stars. The data points
    correspond to the {\em RXTE\/} observations discussed in Psaltis
    et al.\ 1998. The solid and dashed curves are calculated by
    identifying the lower and upper kHz QPOs with the frequencies  
    $\omega_{2-}$ and $\omega_3$, respectively, for two non-rotating
    compact objects with Schwarzchild spacetimes and masses equal to
    1.9 and 2.2~$M_\odot$, respectively. The upper and lower curves   
    for each mass correspond to accretion disk models with scale
    heights $h/r=0$ and $h/r=0.2$, respectively, and provide an
    estimate of the dependence of the predicted frequency on the
    properties of the accretion disk.}
\end{figure}

Figure~3 shows the observed correlation between the lower and upper
kHz QPOs in a number of neutron-star systems (after Psaltis et al.\ 
1998) and compares it to the predicted relation between the resonant
frequencies $\omega_{2-}$ and $\omega_3$. The predicted relation has a
very weak dependence on the compact object spin and we therefore set
$a_*=0$ for the curves plotted. On the other hand, we allow for
different values of the compact object mass and the scale height of
the inner accretion disk.  We also assume that the maximum possible
Keplerian frequency is that of the innermost stable circular orbit and
this results in an upper bound on the upper kHz QPO frequency
(vertical lines in Fig.\,3; see also Miller et al.\ 1998). We find
that the observed correlation can be accounted for, in the present
picture, if the neutron stars are relatively massive ($\sim
1.9-2.2~M_\odot$), in agreement with Stella et al.\ (1999). Moreover,
small discrepancies between the general relativistic frequencies and
the observed QPO frequencies can be easily accommodated in the present
model (for reasonable values of the disk scale height), because of the
corrections introduced to the mode frequencies by the properties of
the accretion flow.

\begin{figure}[t]
  \centerline{
    \psfig{file=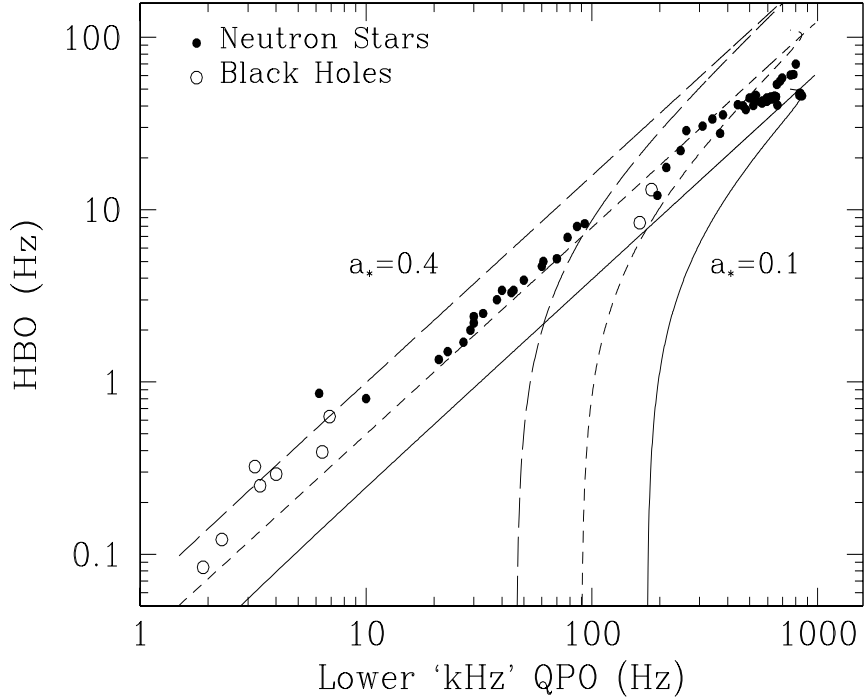,angle=-90,height=10truecm,width=12truecm}}  
  \figcaption[]{\footnotesize Correlation between the frequencies of  
    the HBO and lower kHz QPO (or the peaked noise components
    identified as such) in galactic neutron stars and black holes. The
    data points correspond to the {\em RXTE\/} observations discussed
    in Psaltis et al.\ 1999a. The curves are calculated by identifying
    the HBO and lower kHz QPOs with the frequencies $2\omega_{1}$ and 
    $\omega_{2-}$, respectively, for three compact objects with Kerr
    spacetimes, masses equal to 2.2~$M_\odot$, and different specific 
    angular momenta per unit mass ($a_*=0.1,0.2,0.4$). The upper and
    lower curves for each value of the angular momentum correspond to
    accretion disk models with scale heights $h/r=0$ and $h/r=0.1$,   
    respectively, and provide an estimate of the dependence of the
    predicted frequency on the properties of the accretion disk.}
\end{figure}

Note here that in the absence of any broadening mechanisms, the lower
kHz QPO can be narrower than the upper kHz QPO (compare
equations~[\ref{do1}] and [\ref{do2}]). This is consistent with the
observational fact that, when the two kHz QPOs have different
fractional widths, the lower kHz QPO is systematically narrower than
the upper kHz QPO (see, e.g., M\'endez et al.\ 1998; also van der
Klis, private communication).  Accounting for this property in a
beat-frequency model of the kHz QPOs, in which the lower kHz QPO is
produced by the beat of the upper kHz QPO with a coherent oscillation,
requires the upper kHz QPO peak to be additionally broadened by an
external mechanism.

Figure~4 shows the observed correlation between the HBO and lower kHz
QPO (or the peaked noise component identified as such) in a number of
neutron-star and black-hole systems (after Psaltis et al.\ 1999a) and
compares it to the predicted relation between the resonant frequencies
$2\omega_1$ and $\omega_{2-}$. The predicted relation has a weak
dependence on the mass of the compact object (see also Stella et al.\ 
1999) and we therefore set it equal to $2.2~M_\odot$, allowing for
different values of the compact object spin and disk scale height. We
again find that the observed correlation can be accounted for in the
present picture, if the spins of the compact objects are in the range
$a_*\sim 0.1-0.4$. Note here that, when the hydrodynamic corrections
are non-negligible and hence when $\omega_1\ll\Omega-\Omega_\perp$,
the resonance peaks at $2\omega_1$ are broad and insignificant and
therefore will not appear as prominent QPO peaks (cf.\ Fig\,2).

In the physical mechanism we discuss in this paper, the shape of the
broad-band noise of neutron-star and black-hole systems is the result
of multiplying the intrinsic power spectrum of density perturbations
at the outer disk with the square of the frequency response of the
$(m=0,n=0)$ mode. If the power spectrum of the density perturbations
is relatively flat over a range of frequencies around $\omega_{\rm
  b}$, then the resulting power spectrum will be flat at low
frequencies, have a cut-off at
 \begin{equation}
  \omega_{\rm b}\simeq 10\left(\frac{\alpha}{0.1}\right)
 \left(\frac{\delta r/r}{0.1}\right)^{-1} 
 \left(\frac{h/r}{0.1}\right)^2 
 \left(\frac{\Omega}{1~\mbox{kHz}}\right)~\mbox{Hz}\;,
 \label{breakf}
\end{equation}
and drop above it as $\omega^{-2}$. Assuming that the radial drift
timescale in the transition radius is equal to the viscous timescale,
the relation between the break and HBO frequencies depends mostly on
the fractional width $\delta r/r$ and scale height $h/r$ of the
transition annulus. Modeling, therefore, the correlation discovered by
Wijnands \& van der Klis (1999) between these two variability
components requires a detailed study of the accretion flow properties,
which is beyond the scope of the current paper. According to the
observed correlation $\omega_{\rm b} \sim \omega_1$, and since in our
model $\omega_1 \sim \Omega^2$, this would require 
\begin{equation}
  \left(\frac{\delta r}{r}\right)\left(\frac{h}{r}\right)^{-2}
  \sim \Omega^{-1}\;.
\end{equation}
Note though that the radial drift timescale at the transition radius
cannot be, by construction, strictly equal to the viscous timescale
(even though we assumed it to be for simplicity) and, therefore, the
above correlation is not a strict requirement for our model. In any
case, however, equation~(\ref{breakf}) shows that the overall
properties of the $(m=0,n=0)$ mode are consistent with the shapes and
break frequencies of the noise components observed in both
neutron-star and black-hole systems (see, e.g., Wijnands \& van der
Klis 1999).

We do not attempt here to compare in detail the predictions of our
model to specific sources and in particular to neutron stars for
which the quality of the data is best, because of a number of
systematic effects that introduce uncertainties significantly larger
than those of the frequency measurements. First, in calculating the
QPO frequencies we assumed a Kerr spacetime around the compact object,
which is inconsistent with the high neutron-star spin frequencies
required in this model (see discussion below). Second, the
hydrodynamic corrections to the QPO frequencies depend strongly on the
local temperature and the largely unknown viscosity law and are
specific to the response of the particular physical quantity (i.e.,
the density) we chose to calculate (see, e.g., the discussion in
Bendat \& Pierson 1971, \S2.4). Third, the centroid frequencies of
relatively wide QPO peaks depend not only on the shape of the
response function but on the power spectrum of the perturber, as well.
Finally, for relatively weak QPO peaks, the uncertainties in the
measurement of their centroid frequencies is often dominated by
uncertainties in the subtraction of the noise continuum.

Bearing in mind all these caveats, neglecting the hydrodynamic
corrections, and for a Kerr spacetime and transition radii larger than
the radius of the innermost stable circular orbit, the resonance
frequencies in our model follow the relations (Stella et al.\ 1999)
\begin{equation}
\omega_{2-} \sim \omega_3^{5/3}
\end{equation}
and
\begin{equation}
\omega_{1} \sim \omega_3^2\;.
\end{equation}
The predicted power-law dependences are consistent with the observed
correlations empirically inferred in detailed statistical studies of a
number of neutron-star sources (Fig.\ 2b in Psaltis et al.\ 1998;
Fig.\ 8 in Psaltis et al.\ 1999b). The clear exception appears to be
the correlation between the HBO and upper kHz QPO frequencies in
Sco~X-1, for which the data favor a significantly flatter relation.
It seems, however, likely (see Fig.\,2b in Psaltis et al.\ 1999a) that
the apparent flat correlation in this source is an artifact of
treating as similar data points that follow two distinct branches
(i.e., the `HBO' and `intermediate' branches) of the correlations
discussed in Psaltis et al.\ (1999a). Whether Sco~X-1 exhibits a
behavior similar to that of 4U~1728$-$34 (see Ford \& van der Klis
1998) or not cannot be resolved with the currently available data.

\section{DISCUSSION}

We have studied a mechanism for explaining the broad-band X-ray
variability properties of accreting compact objects, in which a sharp
transition in the accretion disk properties acts as a low-band pass
filter with multiple resonances of the density perturbations produced
outside of it. Similarly to the model proposed by Stella et al.\ 
(1999), all QPO frequencies (which we identify with the various
resonances) are determined mostly by the properties of the metric
exterior to the compact object. As a result, the mechanism we studied
here is applicable to both accreting neutron stars and black holes and
can account naturally for the observed correlations between QPO
frequencies in systems with widely different properties. Moreover, the
same physical mechanism is responsible for the broad-band noise
spectra of such systems and is roughly consistent with their observed
shapes and break frequencies.

Besides the particular modes we have identified with the observed QPOs
in \S4 and their harmonics, a number of additional resonances exist
that could, in principle, produce detectable QPOs at different
frequencies.  First, for the lowest order modes, resonances occur also
at $\omega_{2+}$ and $\omega_{1+}$, which have not been detected as
QPOs in any galactic source.  However, both these frequencies are
significantly larger than the Keplerian frequency at the transition
radius, which corresponds to the fastest timescale there. It is
therefore plausible that the power spectrum of the perturber shows a
sharp cut-off at $\omega_3$ and hence the power at a resonance with
higher frequency is reduced below the detection limits. Moreover,
modes of order ($m$,0) have resonances at frequencies $\omega\simeq
m\Omega\pm\kappa$ with small corrections similar to the ones of
equation~(\ref{o2pm}) and modes of order $(0,n)$ have resonances at
frequencies $\omega\simeq n\Omega_\perp$ with small corrections
similar to the ones of equation~(\ref{o1}). The amplitudes of the
potentially detectable QPOs at these frequencies depend on the
strength of the density perturbations at these modes generated in the
outer accretion disks, which is unknown a priori.  Their predicted
frequencies, however, are particular to the mechanism studied here
(see discussion in Miller 2000) and their detection will provide
additional support to the hypothesis that QPOs occur at fundamental
dynamical frequencies in the accretion disks. It is important to
stress, however, that all the additional resonances occur at
frequencies comparable to or higher than the local Keplerian frequency
$\Omega$. Indeed the only low-frequency QPOs in the current picture
occur at frequencies comparable to $\omega_{1-}$ and its harmonics.

In the model discussed here, the peak separation of the kHz QPOs is
not related to the spin frequency of the neutron star. Instead, it is
nearly equal to the epicyclic frequency of a Keplerian orbit close to
the compact object. This property of the model solves a number of
problems related to the beat-frequency interpretation of kHz QPOs,
such as their variable peak separation (see, e.g., van der Klis et
al.\ 1997; M\'endez et al.\ 1998; see, however, Miller 2000), and of
the nodal-precession interpretation of the HBO, such as the
unphysically large required moments of inertia (see, e.g., Stella \&
Vietri 1998; Psaltis et al.\ 1999b; Kalogera \& Psaltis 2000).
Moreover, it does not require all neutron stars that show kHz QPOs to
be spinning at very similar frequencies (Stella et al.\ 1999), a
result that has important implications for their spin-up and the
possibility of detecting gravitational waves from such systems (see,
e.g., Bildsten 1998; Andersson, Kokkotas, \& Stergioulas 1999).

The correlation between the HBO and upper kHz QPO frequencies in
neutron-star systems depends on their angular momenta and hence allows
us to estimate their spin frequencies, assuming a value for the
stellar moment of inertia $I$ (see, e.g., Psaltis et al.\ 1999b).
Adopting $I\lesssim 10^{45}M$~gr~cm$^3$ (see, e.g., Cook, Shapiro, \&
Teukolsky 1994; also Kalogera \& Psaltis 2000), we find that, in all
sources showing HBO and kHz QPOs, the neutron stars must be spinning
at frequencies $\gtrsim 500$~Hz. Therefore, the burst oscillations,
which are observed at frequencies $\sim 300-350$~Hz, cannot occur at
the stellar spin frequencies. Indeed, it would be too coincidental for
the stellar spin frequency, as inferred from the burst oscillations,
to be as similar to the peak separation of the kHz QPOs as is inferred
observationally for some sources (see, e.g., M\'endez \& van der Klis
1999).

The similarity between the frequencies of the burst oscillations and
the peak separation of kHz QPOs in the four sources in which all these
phenomena have been observed (see, van der Klis 1998) makes tempting
the identification of the burst oscillations with a disk mode that is
excited only when the neutron stars show thermonuclear flashes. In
fact, this disk mode must occur at a frequency comparable to the
maximum epicyclic frequency of a Keplerian orbit near the compact
object and may be related to the trapping of modes studied by Kato
(1990) and Nowak \& Wagoner (1991). (See also the discussion of
$g-$modes in Titarchuk, Lapidus, \& Muslimov 1998). This would
eliminate the problems that arise from the fact that the frequencies
of the burst oscillations are not constant but vary in time (see,
e.g., Strohmayer et al.\ 1996) and that their amplitudes are
non-negligible even in the tails of X-ray bursts (see, e.g.,
Strohmayer et al.\ 1998). Detection of burst oscillations in the
millisecond X-ray pulsar SAX~J1808.4$-$3658 (Wijnands \& van der Klis
1998; Chakrabarty \& Morgan 1998) or of coherent pulsations in any
source showing kHz QPOs will be crucial in assessing these
possibilities.

\acknowledgements

We thank Paolo Coppi, Jean-Pierre Lasota, Wlodek Klu\'zniak, Rashid
Sunyaev, Luigi Stella, and especially Shoji Kato for many useful
discussions on the dynamics of accretion disks. We are grateful to the
organizers of the ITP workshop ``Black Holes in Astrophysics'', during
which this work was initiated and to the participants of the 1999
Aspen Summer Workshop on X-ray Astronomy for many useful discussions.
We thank Deepto Chakrabarty, Fred Lamb, Cole Miller, Luigi Stella, and
Michiel van der Klis for reading critically the manuscript. We also
thank an anonymous referee for comments and suggestions that improved
the paper. D.\,P.\ acknowledges the support of a postdoctoral
fellowship of the Smithsonian Institution. This research was supported
in part by the National Science Foundation under Grant No.\ 
PHY94-07194.

\end{document}